# Nonunitary Classically Stable HD Gravity

Sergio De Filippo and Filippo Maimone

*Dipartimento di Fisica "E.R. Caianiello", Università di Salerno,*
*84081, Baronissi (SA) Italy and I.N.F.M., I,N,F.N, Salerno. defilippo@sa.infn.it*

**Abstract.** Classical instability in fourth order gravity is cured at the expense of unitarity. The appearance of hidden degrees of freedom replicating those of ordinary matter allows for ordinary thermodynamic entropy and black hole entropy to be identified with von Neumann entropy. The emergent picture gives a substantial agreement with B-H entropy and Hawking temperature.

## INTRODUCTION

As a consequence of the so called information loss paradox emerging from black hole physics, "for almost any initial quantum state, one would expect … a nonvanishing probability for evolution from pure states to mixed states" [1]. Although this contravenes a cherished principle of quantum theory, which postulates a unitary time evolution of a state vector in a Hilbert space, the crucial issue is to determine if it necessarily gives rise to a loss of quantum coherence or to violations of energy-momentum conservation too large to be compatible with ordinary laboratory physics. In this respect it is not difficult to construct non-Markovian toy models where rapid information loss occurs at the Planck scale, but negligible deviations from ordinary dynamics occur at laboratory scales [1].

As to the large entropy production in gravitational collapses, it should most naturally occur in the high curvature region, where one expects new physics to emerge, while the event horizon has nothing peculiar for a free falling observer. Of course such a view of Bekenstein-Hawking (B-H) entropy ascribes it to the collapsed matter and, then, it requires a mechanism for the elimination of the singularity. This does not rule out the identification of the entropy of Hawking radiation as coming from the horizon within a local viewpoint. As might be expected, in passing from the the horizon, where quantum field theory in curved space-times is expected to work, to the region close to the classical singularity, we have to pay the price, in the absence of a full theory of quantum gravity, to rely on partially heuristic arguments and some guessing work, which however can be carried out by rather natural assumptions.

On the other hand the issue of B-H entropy is closely connected with that of the microscopic foundations of thermodynamics: "...in order to gain a better understanding of the degrees of freedom responsible for black hole entropy, it will be necessary to achieve a deeper understanding of the notion of entropy itself. Even in flat space-time, there is far from universal agreement as to the meaning of entropy -- particularly in quantum theory -- and as to the nature of the second law of thermodynamics" [2].

In looking for a non-unitary theory accounting for the information loss paradox and avoiding collapse singularities, we will start from higher derivative (HD) gravity. It achieved great popularity since an inflationary solution was obtained without invoking phase transitions in the very early universe, from a field equation containing only geometric terms [3]. A renewed attention was sparked later on by the appearance of HD gravitational terms in the low-energy effective action of string theory and in the holographic renormalization group, as well as by a growing interest in brane worlds within HD gravity [4]. However, although HD theories are natural generalizations of Einstein gravity, already on the classical level they are unstable for the presence of negative energy fields giving rise to runaway solutions. On the quantum level, as to unitarity, an optimistic conclusion is that "the S-matrix will be nearly unitary" [5]. The crucial obstacle in defining HD gravity as a sound physical theory, i.e. the presence of ghosts, seems in fact to point to non-unitarity and to a possible mechanism for avoiding singularities, thanks to short range repulsive terms.

Here a specific non-unitary realization of HD gravity, compatible with the wavelike properties of microscopic particles, as well as with a gravity-induced emergence of classicality, is shown to give strong indications for the elimination of singularities on a trans-Planckian scale. The present setting suggests that B-H entropy may be identified with the von Neumann entropy of the collapsed matter, or equivalently with the entanglement entropy between matter and hidden degrees of freedom, both close to the smoothed singularity. This viewpoint is corroborated by the attractive features of the Newtonian limit of the model, including the possibility of a unified notion, as von Neumann entropy, both for B-H and ordinary entropy.

## CLASSICALLY STABLE FOURTH ORDER GRAVITY

Long ago deWitt [5] and Stelle [6] analyzed the improved ultraviolet behavior of HD gravity theories stemming from cancellations that are "analogous to the Pauli-Villars regularization of other field theories" [6] and are due to the presence of negative energy fields, which in their turn are the source of instability.

A remedy for the ghost problem is proposed below, where, like in Ref. [7], classical instability is cured at the expense of unitarity. Consider the classical action for a fourth order theory of gravity including matter

$$S = S_G[g_{\mu\nu}] + S_{mat}[g_{\mu\nu}, \psi^+, \psi]$$
$$= -\int d^4x \sqrt{-g} \left[ \alpha R_{\mu\nu} R^{\mu\nu} - \beta R^2 + \frac{1}{16\pi G} R \right] + \int d^4x \sqrt{-g} L_{mat} \quad (1)$$

where $L_{mat}$ denotes the matter Lagrangian density. In terms of the contravariant metric density $\sqrt{32\pi G} h^{\mu\nu} = \sqrt{-g} g^{\mu\nu} - \eta^{\mu\nu}$, the Newtonian limit of the static field gives

$$h^{00} \approx 1/r + e^{-\mu_0 r}/(3r) - 4e^{-\mu_2 r}/(3r), \quad (2)$$

where $\mu_0 = [32\pi G(3\beta - \alpha)]^{-1/2}$, $\mu_2 = [16\pi G \alpha]^{-1/2}$ [6].

From Stelle's linearized analysis, the first term in Eq. (2) corresponds to the graviton, the second one to a massive scalar and the third one to a negative energy spin-two field. In fact, in analogy with what is done in Ref. [7] for a toy model, one can introduce a transformation from the field $g_{\mu\nu}$ appearing in the fourth order form of the action to a new metric tensor $\bar{g}_{\mu\nu}$, a massive scalar field $\chi$ dilatonically coupled to the metric and a spin-two massive field $\phi_{\mu\nu}$, this transformation leading to the second order form [8]. To be specific, following Ref. [8] (see Eq. (6.9) apart from the matter term), the action (1) can be rewritten as the sum of the Einstein-Hilbert action $S_{EH}$ for $\bar{g}_{\mu\nu}$, an action $S_{gh}$ for the traceless symmetric ghost field $\phi_{\mu\nu}$ and the scalar field $\chi$ coupled to the metric $\bar{g}_{\mu\nu}$, and a matter action $S_{mat}$, where $g_{\mu\nu}$ is expressed in terms of $\bar{g}_{\mu\nu}$, $\phi_{\mu\nu}$ and $\chi$ (replacing $g_{\mu\nu}$ by $e^{\chi}g_{\mu\nu}$ in Eq. (4.12) in Ref. [8]):

$$S[\bar{g}_{\mu\nu},\phi_{\mu\nu},\chi,\psi^{+},\psi]$$
$$= S_{EH}[\bar{g}_{\mu\nu}] + S_{gh}[\bar{g}_{\mu\nu},\phi_{\mu\nu},\chi] + S_{mat}[g_{\mu\nu}(\bar{g}_{\sigma\tau},\phi_{\sigma\tau},\chi),\psi^{+},\psi]. \quad (3)$$

In $S_{gh}$ above the quadratic part in $\phi_{\mu\nu}$ has the wrong sign [8], which may lead to runaway solutions where both the negative energy of this ghost field and the positive energy of the other degrees of freedom grow exponentially in time. A simple way to get rid of classical instability would be to symmetrize the action with respect to the transformation $\phi_{\mu\nu} \to -\phi_{\mu\nu}$ and to introduce a symmetry constraint on the allowed states with respect to this transformation. This however would eliminate the corresponding repulsive term in Eq. (2), which is a possible candidate in avoiding the singularity in gravitational collapse. Once one accepts non-unitarity, it is rather natural to assume that one can cure the instability, while keeping the short-range repulsive term, by introducing hidden degrees of freedom, i.e. from a quantum viewpoint to accept that the operator algebra involved in defining the dynamics is larger than the observable algebra. To be specific, we double the matter algebra by taking a meta-matter algebra as the product of two copies of the matter algebra, respectively generated by the operators $\psi^{+},\psi$ and $\tilde{\psi}^{+}\tilde{\psi}$. We then define the symmetrized action [9,10]

$$S_{Sym} = \frac{1}{2}\{S[\bar{g}_{\mu\nu},\phi_{\mu\nu},\chi,\psi^{+},\psi] + S[\bar{g}_{\mu\nu},-\phi_{\mu\nu},-\chi,\tilde{\psi}^{+},\tilde{\psi}]\}, \quad (4)$$

which is symmetric under the transformation

$$\bar{g}_{\mu\nu} \to \bar{g}_{\mu\nu},\ \phi_{\mu\nu} \to -\phi_{\mu\nu},\ \chi \to -\chi,\ \psi \to \tilde{\psi},\ \tilde{\psi} \to \psi, \quad (5)$$

If the state space is restricted to those $|\Psi\rangle$ that are generated from the vacuum by symmetrical operators, i.e. invariant with respect to this transformation, then $\forall F$

$$\langle\Psi|\{F[\bar{g}_{\mu\nu},\phi_{\mu\nu},\chi,\psi^+,\psi,\tilde{\psi}^+,\tilde{\psi}]-F[\bar{g}_{\mu\nu},-\phi_{\mu\nu},-\chi,\tilde{\psi}^+,\tilde{\psi},\psi^+,\psi]\}|\Psi\rangle=0. \quad (6)$$

This implies that, as usual with constrained theories, allowed states do not give a faithful representation of the original algebra, which is then larger than the observable algebra. In particular the constrained state space cannot distinguish between $F[\psi^+,\psi]$ and $F[\tilde{\psi}^+,\tilde{\psi}]$, by which the $\tilde{\psi}$ operators are referred to hidden degrees of freedom, according to a standard terminology for non-unitary models [1], while only the $\psi$ operators represent matter degrees of freedom. On a classical level the constraint implies that $\psi$ and $\tilde{\psi}$ are to be identified while the $\chi$ and $\phi_{\mu\nu}$ fields vanish and, as a consequence, the classical constrained action is that of ordinary matter coupled to ordinary gravity:

$$S_{Cl}[\bar{g}_{\mu\nu}\psi^+,\psi]=S_{EH}[\bar{g}_{\mu\nu}]+S_{mat}[\bar{g}_{\mu\nu},\psi^+,\psi], \quad (7)$$

as $S_{gh}[\bar{g}_{\mu\nu},0,0]=0$ (Eq. (6.9) in Ref. [6]) and $g_{\mu\nu}(\bar{g}_{\sigma\tau},0,0)=\bar{g}_{\mu\nu}$ (Eq. (4.12) in Ref. [6] with $e^\chi g_{\mu\nu}$ replacing $g_{\mu\nu}$). While this modification of fourth order gravity is expected to affect its ultraviolet behavior, still it does not worsen it for one-loop meta-matter to meta-matter amplitudes, at variance with the trivial symmetrization of the original theory. It should also be remarked that one could limit symmetrization to the ghost field, without involving the scalar field, especially if one were interested in the cosmological implications of keeping the dilaton field in the classical action.

A final remark is in order as to the possible rereading of the present model as a bimetric HD theory where two worlds interact only by means of a coupling between the metrics (of the fourth order formalism) [11]. In fact the model can be defined by replacing $\bar{g}_{\mu\nu}$, $-\phi_{\mu\nu}$ and $-\chi$ in the second term of the right hand side of Eq. (4) by three independent fields and adding an interaction, including Lagrange multipliers, leading to the necessary identifications. However we do not commit ourselves to the prevailing view pointing to underlying higher dimensional theories, even though a natural setting could appear to be an extension of the Randall-Sundrum model, with two positive tension flat branes separated by one intermediate negative tension flat brane [12], where $\psi$ and $\tilde{\psi}$ meta-matters reside on distinct positive tension branes.

It can be shown [13, 14, 10] that the Newtonian limit of the model, while compatible with the wavelike properties of microscopic particles, gives rise to localization and decoherence for the center of mass motion of macroscopic bodies. In particular it leads to a sharp localization threshold around $10^{11}$ proton masses. While the symmetry constraint avoids that hidden degrees of freedom may be "available as either a net source or a sink of energy" [1], only the generator of the unitary dynamics in the enlarged algebra is strictly conserved. The physical energy undergo fluctuations, which, though irrelevant on a macroscopic scale, may lead, starting for instance from an eigenstate of the physical Hamiltonian, to a microcanonical mixed state.

# BLACK HOLE ENTROPY

While the Newtonian limit is a well defined nonrelativistic model with unambiguous consequences for ordinary laboratory physics, which in principle can be either confirmed or disproved experimentally, the present application to black holes is a rather bold extrapolation. The soundness of its results and its rather natural assumptions make us confident that it may contain some elements of a future consistent theory. To evaluate within our model the linear dimension of a collapsed matter lump of mass $M$, replacing the classical singularity, we use the result that localized solutions with superposed observable and hidden matter are present as far as the localization length $\Lambda \equiv (2\hbar^2 R^3/GM^3)^{1/4}$ is fairly smaller than the lump radius $R$ [13, 14, 10]. Since we are now considering the trans-Planckian limit, we may put $\mu_0 = \mu_2 = 0$, by which in this limit gravity acts only between observable and hidden meta-matter and then it is ineffective in collapsing matter below the delocalization threshold. This gives for the radius of the collapsed lump [9]

$$R = \hbar^2/(GM^3). \tag{8}$$

If in the region beyond the horizon we put $x = v - \int dr [1 - 2MG/(rc^2)]^{-1}$, the metric becomes

$$ds^2 = [1 - 2MG/(rc^2)]^{-1} dr^2 - [1 - 2MG/(rc^2)] dx^2 + r^2 [d\theta^2 + \sin^2\theta d\phi^2], \tag{9}$$

where $(v, r, \theta, \varphi)$ denote the ingoing Eddington-Finkelstein coordinates [15]. If we trust (8) as the minimal length involved in the collapse, we are led to assume that a full theory of quantum gravity should include a mechanism regularizing the singularity at $r = 0$ by means of that minimal length. In particular it is natural to assume that the $x$ width of the collapsed matter lump is $\Delta x = R$. On the other hand we are proceeding according to the usual assumption, or fiction, of QM on the existence of a global time variable, at least in the region swept by the lump. In fact the most natural way to regularize (9) is to consider it as an approximation for $r > R$ of a regular metric, whose coefficients as $r \to 0$ correspond to the ones in (9) with $r = R$, in which case there is no obstruction in extending the metric to $r < 0$, where taking constant coefficients makes $\partial/\partial r$ a time-like Killing vector. As a consequence, the relevant space metric in the region swept by the collapsed lump is [9]

$$ds^2_{SPACE} \approx [2MG/(Rc^2)] dx^2 + R^2 [d\theta^2 + \sin^2\theta d\phi^2] \tag{10}$$

and the volume of the collapsed lump is then $V \approx R^2 \Delta x \sqrt{MG/(Rc^2)} = \hbar^5 M^{-7}/(G^2 c)$.

At equilibrium one has a microcanonical ensemble with energy

$$E = Mc^2 + GM^2/R \cong G^2 M^5/\hbar^2, \text{ if } M \gg M_P, \tag{11}$$

and energy density $\varepsilon = G^4 c M^{12}/\hbar^7$, where $M_P = \sqrt{\hbar c/G}$ is the Planck mass.

Since this energy density corresponds to a very high temperature, not to be mistaken for the Hawking temperature, the matter can be represented by massless fields, whose equilibrium entropy is given by

$$S \approx \left(K_B/\hbar^{3/4} c^{3/4}\right) \varepsilon^{3/4} V = K_B G M^2/(\hbar c). \tag{12}$$

Of course this result can be trusted at most for its order of magnitude, the uncertainty in the number of species being just one part of an unknown numerical factor. With this proviso, common to other approaches, Eq. (12) agrees with B-H entropy [2].

Finally, if we give for granted that a future theory of quantum gravity will account for black hole evaporation, we can connect the temperature $T \approx \sqrt[4]{\varepsilon \hbar^3 c^3}/K_B \approx c G M^3/K_B \hbar$ of the collapsed lump with the (spectral) temperature of the radiation at infinity. If we model radiation by massless fields, emitted for simplicity at a constant temperature, this temperature is defined in terms of he ratio $E_\infty/S_\infty$ of its energy and its entropy. It is natural to assume that, "once" thermodynamic equilibrium is reached due to the highly non-unitary dynamics close to the classical singularity, no entropy production occurs during evaporation, by which $S_\infty = S$. Then, if $Mc^2$ is the energy of the total Hawking radiation spread over a large space volume, its temperature agrees with Hawking temperature, i.e. $T_\infty = (E_\infty/E)T \approx c^3 \hbar/MGK_B$ [2].